\documentclass[letterpaper,twocolumn,10pt]{article}
\usepackage{usenix,epsfig,endnotes,rotating,footnote}
\usepackage{graphicx}
\usepackage{amsmath}
\usepackage{subcaption}

\begin{document}

\date{}

\title{\Large\bf Chill-Pass: Using Neuro-Physiological Responses to Chill Music to Defeat Coercion Attacks%
	}

\author{
{\rm Max Wolotsky}\\
Cal Poly Pomona
\and
{\rm Mohammad Husain}\\
Cal Poly Pomona
\and
{\rm Elisha Choe}\\
Sandia National Laboratories
\thanks{Sandia National Laboratories is a multi-program laboratory managed and operated by Sandia Corporation, a wholly owned subsidiary of Lockheed Martin Corporation, for the U.S. Department of Energy’s National Nuclear Security Administration under contract DE-AC04-94AL85000.}
} 

\maketitle


\subsection*{Abstract}

Current alphanumeric and biometric authentication systems cannot withstand situations where a user is coerced into releasing their authentication materials under hostile circumstances. Existing approaches of coercion resistant authentication systems (CRAS) propose authentication factors such as implicit learning tasks, which are non-transferable, but still have the drawback that an attacker can force the victim (causing stress) to perform the task in order to gain unauthorized access. Alternatively, there could be cases where the user could claim that they were coerced into giving up the authentication materials, whereas in reality they acted as an insider attacker. Therefore, being able to detect stress during authentication also helps to achieve non-repudiation in such cases. To address these concerns, we need CRAS that have both the non-transferable property as well as a mechanism to detect stress related to coercion. In this paper, we study the feasibility of using Chill (intensely pleasurable) music as a stimulus to elicit unique neuro-physiological responses that can be used as an authenticating factor for CRAS. Chill music and stress are both stimuli for a neuro-chemical called Dopamine. However, they release  the Dopamine at different parts of the brain, resulting in different neuro-physiological responses, which gives us both the non-transferable and stress-detection properties necessary for CRAS. We have experimentally validated our proposed Chill music based CRAS using human subjects and measuring their neuro-physiological responses on our prototype system. Based on the 100 samples collected from the subjects, we were able to successfully authenticate the subjects with an accuracy of over 90\%. Our work not only demonstrates the potential of Chill music as a unique stimulus for CRAS, but also paves the path of wider adoption of CRAS in general. 

\begin{figure*}[!t]
	\begin{center}
		\includegraphics[width=.8\textwidth]{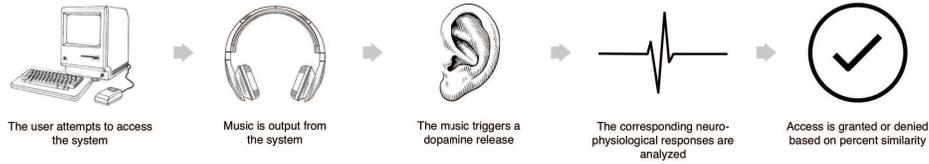}
	\end{center}
	\caption{\textit{After a user's neuro-physiological responses have been registered into the proposed authentication system, authentication can be completed by having a user listen to their selected Chill music.}}
	\label{fig:demo}
\end{figure*}

\section{Introduction}
Imagine a hypothetical user Alice who tries to practice good security principles. Alice has a long and random alphanumeric password and supplements her authentications using multiple factors, one of which might also include using her biometric features such as fingerprint. 
A desperate attacker, may however, instead of trying to brute force an impossibly long and random password or acquire other authentication factors, resort to social-engineering, drugging, or threatening Alice. The problem exists that no matter how strong the password is, or how many extra steps one takes in order to protect the authentication information, current alphanumeric and biometric authentication systems cannot defend against the aforementioned attack vectors that we refer to as coercion attacks~\cite{gupta2013exploiting}. \\
\indent However, there are a plethora of security critical systems defending financial systems and national security related installations, which cannot afford to risk such situations where a valid user is physically compromised or coerced, and their authentication materials are transferred to an attacker. Initial proposals to address coercion attacks are based on generating innovative authentication materials using implicitly learned tasks (i.e. habitual bias in performing a certain task), which could not be consciously explained to an attacker~\cite{bojinov2012neuroscience,diwakar2015theme} and therefore  prevented \textit{authentication materials from being transferred} during a coercion attack. However, these proposals did not explicitly cover the second (and more realistic) scenario of coercion, where an attacker forces the victim \textit{to perform the authentication process} for them so that the attacker can gain unauthorized access to the system. Alternatively, there could be cases where the user could claim that they were coerced into give up the authentication materials, whereas in reality they acted as an insider attacker. In such cases, being able to detect stress during authentication helps to achieve non-repudiation as well. In general, this form of coercion is more relevant to biometrics because although in some cases it is shown possible to copy a biometric feature~\cite{wiehe2004attacking}, it is not always trivial to do so, and it is often easier to just force the victim into authenticating and giving control of the authenticated system to the attacker. Preventing this kind of coercion attack is much more difficult, and requires an approach that could factor in some mechanism to identify when a user is under high levels of stress (which could possibly be due to coercion). \\
\indent A common approach to study stress is neuro-physiological measurements, which led us to the unique neuro-chemical associated with it: \textit{Dopamine}. Interestingly enough, in the domain of neuro-physiology, another widely studied stimulus of Dopamine release is Chill music~\cite{salimpoor2011anatomically}. When a person listens to a music that he/she find intensely pleasurable, it causes a certain amount of Dopamine to be released in their brain (but in a different region than stress), which also leads to a noticeable change in their neuro-physiological responses such as heart rate, skin conductance, and brain wave to name a few. In addition, the location and effect of these two types of Dopamine release are distinct: Chill music related Dopamine is found in mesolimbic region whereas stress related Dopamine is found in ventral straitum area~\cite{StressDopamine}. So, effectively if we could use the neuro-physiological responses (due to Dopamine release) to Chill music as an authenticating factor, then we could also detect stress due to coercion, as the neuro-physiological responses will be distorted due to the effect of a stress related Dopamine release~\cite{everly2002anatomy,thayer2012meta}. Thus, CRAS based on Chill music would not only meet the non-transferable criterion but also detect stress and thwart a wider range of coercion attacks. \\
\indent In this paper, we have experimentally validated our proposed Chill music based CRAS using human subjects by trying to answer the following fundamental questions:

\begin{itemize}
	\item Does a human subject elicit similar neuro-physiological response to a given Chill music over multiple trials?
	\begin{itemize}
		\item If yes, how consistent is it over time?
	\end{itemize}
	\item Are the responses of two human subjects to each other's Chill music distinctly identifiable?
	\begin{itemize}
		\item How about their responses to random pieces of music?
	\end{itemize}
\end{itemize}

We have implemented a prototype of the system using neurological and physiological sensors and collected data from human subjects\footnote{Appropriate human subject research approval was received from the Institutional Review Board} to answer the aforementioned questions and validate our proposed CRAS. A simple diagram showing the basic overview of our prototype system is shown in Figure \ref{fig:demo}. \\
\indent In the remaining sections of the paper, we discuss background information about our prototype CRAS (Section 2), the methods by which we conducted our experiments (Section 3), results and analysis gathered from our experiments (Section 4), a discussion of how our prototype system applies to real-world use cases (Section 5), an overview of previous works done in this domain (Section 6), and our conclusions based on our findings (Section 7).

\section{Background Information}
An important quality of our prototype CRAS is that the neuro-physiological responses used as authentication materials should stay relatively consistent over time, and the user's baseline neuro-physiological responses should not reveal the responses used as authentication materials. To verify that our prototype will have these qualities, we were required to make important design choices, specifically when dealing with the music selection and registration processes. 


\subsection{Music}
We have chosen music as the stimulus for our prototype CRAS because it can provide a guarantee that neuro-physiological responses will be different during authentication than they are during baseline, and will also be easily reproducible when required for authentication. However, in order to use neuro-physiological responses to music effectively, it is necessary that we limit the music that we select for use in authentication to be a subset of music that contains certain desirable properties.



\subsubsection{The Chill Effect}
A Dopamine release experienced by the stimulation of highly pleasurable music is often referred to as the Chill effect~\cite{altenmuller2013strong}. This is because the most common description of such a reaction is feeling a ``chill'' or ``shivers down the spine''~\cite{grewe2009chill}. During the Chill effect, neuro-physiological responses are noticeably altered from their baseline, and can be used to provide a biometric signature for the user experiencing that Dopamine release. \\
\indent Not all pieces of music stimulate the Chill effect when listened to, and two users listening to the same piece of music do not necessarily experience the same change in neuro-physiological responses~\cite{salimpoor2009rewarding}. The occurrence of the Chill effect when listening to a piece of music, and the degree in which neuro-physiological responses change during a Dopamine release are also dependent on both the user, and the piece of music listened to. A piece of music that can stimulate the Chill effect in a user is referred to as Chill music for that user.\\
\indent Instead of using entire pieces of Chill music as stimulus for our prototype CRAS, we chose to use Chill segments in order to significantly reduce the amount of time it takes to conduct authentication. Chill segments are a subset of Chill music that can elicit a Chill response without the entire piece of music being played. Previous studies have shown that certain segments of music are more likely than others to elicit a Chill response, and given several subjects that listen to the same piece of music and experience a Chill response, a majority of the Chill responses occur at similar time periods~\cite{nagel2008psychoacoustical}. Because of this property, we can localize a Chill response to a certain segment of music, and avoid playing the entire piece of music. \\

\subsubsection{Music Selection}
In order to obtain optimal results during authentication, we impose certain rules on the music that is chosen during the music selection process, leaving the user with a subset of pieces of music which are optimal for use in a Chill music based CRAS. The criteria we impose on music selection are that the music must be Chill music for the user, the music must be non-lyrical, and the music should not be linked to a specifically volatile memory of the user (a song played on a first date for example). \\
\indent The issue we are trying to avoid by using non-lyrical music is that lyrics can be commonly linked to volatile memories, and the emotions related to such memories can possibly cause neuro-physiological responses to vary depending on the current emotion the user is feeling. This issue is referred to as day dependence, and is one of the main reasons why the perceived pleasantness of a piece of music can deteriorate over time. There are many other reasons why day dependence may occur, but we have found that eliminating lyrics from the music used in our prototype CRAS is the most effective way to mitigate this issue. Although the issues associated with day dependence aren't always significant, we find that by avoiding the use of music that can easily be subject to day dependence that we can significantly enhance the accuracy of our authentication system. We explore issues associated with day dependence in more detail in our experiment.

\subsection{Registering Into The System}
\begin{figure}[t]
	\begin{center}
		\includegraphics[width=.5\textwidth]{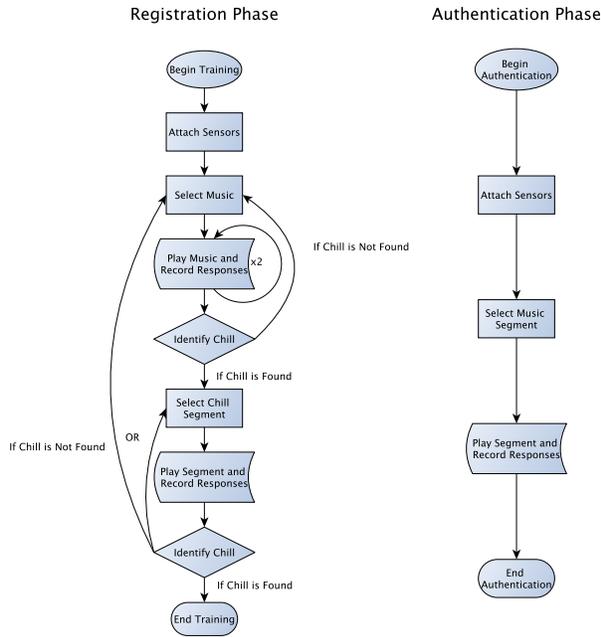}
	\end{center}
	\caption{\textit{For a user to register into the authentication system, they must choose music that stimulates a Chill response. After they have determined music that stimulates this response, a one-minute segment which stimulates a Chill response must be selected. Once this one-minute segment is determined, the user can be registered into the system. The authentication phase can be conducted by choosing a music segment that has been registered to the expected user and playing it to the authenticating user while monitoring their neuro-physiological responses.}}
	\label{fig:taphase}
	
	
\end{figure}

Before being able to authenticate into the prototype CRAS, it is required that a user register their neuro-physiological responses to Chill music with the system. The registration phase consists of the prototype system storing important information to be used during future authentication attempts, and consists of selecting music to be used for authentication, truncating the selected music into a one-minute segment, and finally storing the neuro-physiological responses collected during the stimulation of the selected music segment. Once the desired neuro-physiological responses are collected, they can be used for comparison in future authentication attempts. This process is conducted in order to assure that the neuro-physiological responses used for authentication will be as consistent over time as possible. An overview of the registration phase, and the subsequent authentication phase, can be seen in Figure~\ref{fig:taphase}.

\subsubsection{Verifying Music as Chill Music}
Before the registration phase begins, the user should select multiple pieces of music that they perceive to be highly pleasurable, and which follow the criteria highlighted in previous sections, such as being non-lyrical. A piece of music can then be selected from the set the user has chosen, and the user's neuro-physiological responses to that music can be collected. If the collected responses show that the music did not stimulate a Chill response, or failed to produce consistent results, the music and results will be discarded and the user prompted to pick another piece of music. These results are discarded in order to prevent the user from choosing a piece of music that they will not respond consistently to in the future. This process can continue multiple times until one or more pieces of music are identified as Chill music. The advantage of having multiple pieces of music registered is that it will allow for neuro-physiological responses to multiple pieces of music to be used for authentication, providing a heightened level of security and accuracy.

\subsubsection{Verifying a Chill Segment}
After a piece of music chosen by a user is verified as Chill music, the music must go through one further step of being truncated to a one-minute segment. This one-minute segment is used to highlight the area of the music where the Chill response occurs, and reduce the amount of time it takes to authenticate in the future. This one-minute segment can be identified by finding the most significant overlaps in Chill responses from the multiple sets of neuro-physiological responses collected during the initial music verification portion. Once the one-minute segment is determined, the segment goes through the music verification step once more to verify that it is a valid Chill segment, and if it succeeds then the selected segment is registered into the authentication system along with the corresponding neuro-physiological responses. If the segment fails the verification step, another one-minute segment can be tested and the verification step repeated, or a new piece of music can be chosen.

\subsection{Preventing Baseline Attacks}
Our prototype uses Chill music in order to provide us with a stimuli based authentication method. Stimuli based authentication refers to biometric signatures that are only present when the user is introduced to a certain stimulus~\cite{zuquete2010biometric}. For example, unlike a fingerprint, the heart rate of a person changes during various activities. If a person is running at a certain pace, their heart rate differs from their normal pattern, but nonetheless has a pattern of its own~\cite{thayer2012meta}. By using this principle, we can protect biometrics from being stolen during activities not necessary for authentication by using a biometric that is only stimulated during the time of authentication. Adding onto the previous example, if the user's heart rate while running at a certain pace was necessary for authentication, then it could not be stolen from the user in situations where they were not running at that pace. For use in CRAS, because the required neuro-physiological responses are not available at a baseline state, as long as the stimulation required for authentication cannot be achieved while coerced, then the required neuro-physiological responses cannot be stolen from a victim during a coercion attack. \\
\indent Showing that the neuro-physiological responses essential for authentication are only available when introduced to the selected stimulus is important for defending against attacks where the user's baseline neuro-physiological responses are compromised, but it does not account for situations where the user is coerced and subsequently introduced to the selected stimulus. A CRAS can defend against this scenario by using Chill music as the selected stimulus, which will cause a Dopamine release in the user during authentication. If the user does not experience the necessary Dopamine release, or there are external factors affecting the user's neuro-physiological responses to an extent that it prevents the desired responses from occurring, then authentication can be easily denied. In the case of a coercion attack, we expect that the user's neuro-physiological responses will be affected in an extreme way, and therefore would prevent the user from producing accurate authentication materials. Other attacks such as brute force attacks are also ineffective against this type of CRAS, because the system can only perform one authentication per minute (at best). This makes brute force attacks infeasible as they generally require a large amount of attempts to be tried in a short period of time.


%
%


\section{Experimental Procedure}
In this experiment we simulated our prototype CRAS by conducting the registration phase on multiple subjects, and having each subject attempt to authenticate as themselves, as well as all other subjects. We accomplished this by allowing each subject to select a piece of Chill music to be used to register them into the authentication system, and then played them their selected Chill music, other subject's selected Chill music, and several other music samples. These collected responses were later compared to the registered responses for each subject in order to verify that only pairs of responses collected from the same subject would be sufficiently similar. To ensure that all subjects were treated ethically during the duration of this experiment, we received IRB approval to conduct all tests.

\begin{table*}[]
	\centering
	\resizebox{\textwidth}{!}{%
		\begin{tabular}{|c|c|c|c|c|}
			\hline
			\textbf{Subject Number} & \textbf{Selected Chill Music}         & \textbf{Others Selected Chill Music}  & \textbf{Random Constant} & \textbf{Random Variable}                          \\ \hline
			1                       & The City Gates - The Elder Scroll V   & Das Malefitz - Faunts                 & Rain - Joe Hisaishi      & Minas Tirith - The Return of the Kings Soundtrack \\ \hline
			2                       & Aquarium - Nosaj Thing                & The City Gates - The Elder Scroll V   & Rain - Joe Hisaishi      & River Flows in You - Yiruma                       \\ \hline
			3                       & Das Malefitz - Faunts                 & Aquarium - Nosaj Thing                & Rain - Joe Hisaishi      & Full Focus - Armin Van Buuren                     \\ \hline
			4                       & Gerudo Valley - Zelda Ocarina of Time & Das Malefitz - Faunts                 & Rain - Joe Hisaishi      & Moonlight Sonata - Beethoven                      \\ \hline
			5                       & Kinder Blumen - Real Estate           & Gerudo Valley - Zelda Ocarina of Time & Rain - Joe Hisaishi      & Fur Elise - Beethoven                             \\ \hline
		\end{tabular}
	}
	\caption{\textit{Each subject listened to four different pieces of music. The first was a piece of music they selected that stimulated a Chill response. The next was a piece of music that stimulated a Chill response in another subject. The last two pieces were a random constant that was chosen to be the same for all subjects, and a random variable which was chosen to be a piece that no other subject had listened to.}}
	\label{tab:musiclist}
\end{table*}
%

\subsection{Experimental Setup}
All subjects monitored were students between the ages of 18-25, and identified at least three pieces of music that they significantly liked before showing up for the experiment. Of the five subjects monitored in the preliminary experiment, two were male and three female. All music chosen was non-lyrical, and instrumental music was not chosen if the original piece of music contained lyrics. Subjects were instructed to follow music selection criteria highlighted in Section 2. The music played to each subject is shown in Table~\ref{tab:musiclist}. \\
\indent For this experiment we used two different devices to monitor the neuro-physiological responses of our subjects. The NeuroSky Mindwave Mobile was attached to the forehead and ear of each subject and was used in order to monitor Alpha, Beta, Gamma, Delta, and Theta brain waves. The subject was also attached with a MIO Alpha 2 heart rate monitor on their wrist, which was used to sample their heart rate. The NeuroSky headset monitored approximately one neurological response (one for every wavelength) every second, while the MIO Alpha watch monitored about two responses per second.

\subsection{Experimental Method}
Before the experiment began, all subjects were given an initial survey which required them to pre-select three pieces of non-lyrical music that they identified as highly pleasant, and followed the criteria highlighted previously. A single piece of music that the subject pre-selected was chosen in order to conduct the registration phase with. If the piece of music chosen failed to generate valid authentication materials, then another piece of music the subject pre-selected was chosen, and the process repeated until valid authentication materials were obtained. Only one piece of music was registered to each subject in this way. Once the subject registered their neuro-physiological responses using their selected piece of music (and the corresponding one-minute segment for that piece of music), the Chill segment they registered was stored for future use, and they were played multiple segments of other pieces of music. \\
\indent Each subject was played their own registered music segment, the registered music segment of another subject, a constant piece of music played to all subjects, and a random piece of music that no other subject was played. Subjects were given breaks in between listening to each piece of music in order to verify that their responses from listening to the previous piece of music had returned to a baseline level and would not significantly interfere with their responses to the next piece of music. \\
\indent Subjects were played their own registered music segment in order to determine the consistency of their responses to their selected Chill segment. They were played the Chill segment of another user in order to determine if the responses of two subjects to the same piece of Chill music were distinctly identifiable. The constant piece of music and random piece of music were used in order to add as much variety in responses as possible in order to determine if a subjects responses to a random piece of music would match with another subjects registered responses to Chill music.

\subsection{Neuro-Physiological Responses}
When attempting to authenticate a user, we use both their physiological and neurological responses to their selected music in order to identify them. For our purposes, we select specific physiological and neurological responses to monitor based on their relevance to our desired CRAS, and the availability of equipment at our disposal.

\subsubsection{Physiological Responses}
We have chosen heart rate as the main physiological factor for our authentication system for multiple reasons. The first reason we chose heart rate is because of the high-level of convenience associated with monitoring it. Above being easy to monitor, a user's heart rate is relatively unique to that user, and during a Dopamine release the user's heart rate is noticeably affected in a significant and unique way. Heart rate also changes very quickly when the user is exposed to high levels of stress associated with coercion~\cite{everly2002anatomy}. Because of this change, if the user were to be coerced while trying to authenticate, it would be easy to identify the coercion due to their irregular heart rate. \\
\indent Although we use heart rate for the purpose of our experimentation, it is not the only physiological response that would be suitable for a CRAS. Through previous works (which will be discussed in detail later) we have seen that physiological responses including skin conductance, skin temperature, blood pressure, and respiration rate can also be effectively combined to provide authentication~\cite{grewe2007listening,koelstra2012deap}. Skin conductance (like heart rate) is also changed very quickly when high levels of stress are introduced, so it is similarly effective in identifying coercion~\cite{gupta2010fighting}.

\subsubsection{Neurological Responses}
Unlike with physiological responses, we did not necessarily know before experimentation which neurological responses would be most relevant in identifying the Chill effect in a user. Because of this, we used all of the available neurological responses our equipment would allow. These responses included Alpha, Beta, Gamma, Delta, and Theta waves. Although these were the only neurological responses we were capable of monitoring, it is possible that other kinds of neurological responses can also be useful in our proposed CRAS~\cite{armstrong2015brainprint}.

\subsection{Monitoring the Responses}
Each subject went through a minimum of seven monitoring cycles during the duration of the experiment. At minimum, a subject listened to their selected Chill music twice, the corresponding Chill segment twice, another subject's Chill segment once, a constant piece of music played to all subjects once, and a random piece of music played to no other subject once. In many cases subjects went through more monitoring cycles, due to either failure to determine a Chill during the registration phase, or equipment malfunction during the authentication phase. \\
\indent Monitoring cycles were all conducted similarly regardless of whether the responses being monitored were meant for authentication or registration. The only difference during the monitoring cycles for the registration phase was that they were longer in the case where the Chill segment had yet to be determined. Each monitoring cycle began with an equipment test to verify that the neuro-physiological monitoring devices were functioning properly. If any errors were detected in functionality, the devices were reset and the equipment test was repeated until the equipment was verified as working properly. Once the equipment had been tested, a one-minute baseline was taken in order to verify that the subject was in a comfortable state (their heart rate was not highly elevated or lowered), and in order to determine the subject's normal heart rate so that a Chill could be identified. After the one-minute baseline was completed, we began playing music to the subject. \\
\indent Subjects were instructed to focus on the music while it was playing, and were not allowed to interact with smart phones or other electronic devices. They were allowed to close their eyes if they wished, but were permitted to keep their eyes open granted that they did not stare at a single object during the monitoring. This was done to prevent neurological responses from being compromised due to intensely focusing on a single object instead of the music being played. The prohibition of smart phones and other electronics was also to ensure that neurological responses weren't compromised by distracting the subject from the music they were listening to.

\subsection{Determining Chill Music and Chill Segments}
When determining whether or not a piece of selected music can serve as Chill music for a subject, we first determine instances of Chills in the physiological responses of the subject. We look for Chills in the physiological responses because we have seen in previous work that heart rate responds quickly and noticeably to Chills~\cite{everly2002anatomy,thayer2012meta}. We define a Chill mathematically as any range of responses where every point in the range is greater than one standard deviation away from the average of the collected baseline responses. We chose the minimum acceptable range length to be five seconds, so any range of responses less than five seconds in length were not considered to be a Chill. \\
\indent A piece of music was only considered Chill music if a Chill was found in multiple occurrences of listening to that piece of music. In this experiment, after listening to the full piece of selected music twice, the collected responses were checked to see if they both contained a Chill. If both collected responses contained a Chill, then the music was classified as Chill music for the subject and could be used for authentication. Otherwise, a new piece of music had to be chosen and the process repeated. \\
\indent Even if a piece of music was classified as Chill music, it was necessary to determine a Chill segment for that piece of music (a one-minute segment that elicited a Chill response in the subject). There were multiple options for determining this Chill segment, it could either be chosen manually by the user or it could be chosen by an analysis program. In either case, the Chill segment had to be tested to verify that it elicited a Chill response before it could be used for authentication. \\
\indent Most subjects chose to allow the analysis program to determine the Chill segment for them. The analysis program worked by parsing the collected responses from subjects listening to the full Chill music, and comparing the segments of the music where Chill responses occurred in each sample. The analysis program then determined the overlapping segments and output the longest overlapping segment found. 30 seconds before and after the middle of the selected segment was then chosen as the one-minute Chill segment. If the computer generated segment did not work, or the subject preferred to manually identify the Chill segment they wanted to use, 30 seconds before and after the point in the music they chose became their Chill segment. \\
\indent After deciding what Chill segment should be used, we then had to verify that the selected Chill segment elicited a Chill response in the subject. To do this, we simply played the segment to the subject and collected their neuro-physiological responses. If we analyzed a Chill in the responses, then the Chill segment was acceptable, and their collected responses were registered into the authentication system. Otherwise there were two options, either the subject could select a different Chill segment, or they could select a different piece of music and repeat the registration process.

\subsection{Collecting Authentication Attempts}
For this experiment we separated the collected neuro-physiological responses into authentication attempts. An authentication attempt was a pair of neuro-physiological responses from two different monitoring cycles, and included one set of registered responses and one set of authentication responses. All authentication attempts we chose included registered responses that were actually registered to the subjects (we only chose music segments that successfully completed the registration phase). The authentication responses that we compared to the registered responses consisted of all other responses (all collected music segments not used during registration) in order to simulate as many different scenarios as possible. \\
\indent There were four categories authentication attempts could fall into in this experiment, and each category reflected the relationship between the registered responses and the authentication responses. Each attempt was either a comparison of the same subject as in the registered response listening to the same piece of music, the same subject listening to a different piece of music, a different subject listening to the same piece of music, or a different subject listening to a different piece of music. We analyzed and compared results from these categories to show that authentication attempts comparing responses from the same subject could be reliably differentiated from those from different subjects.\\
\indent When comparing collected responses from authentication attempts, the responses were first sanitized in order to verify that the most relevant portions of the responses were compared. First, all baseline was stripped off of the responses so only responses collected while the user was listening to music were compared. Then, the first and last 10 seconds of the responses were removed in order to ensure that responses were not affected by factors other than the user's reactions to the music. This system highlighted the 40-second middle portion of the music segment, which due to the segment selection process was significantly more likely to contain a Chill response than the first or last 10 seconds.

\subsection{Time Lapse Experiment}
We selected two subjects to come back after a variable amount of time after the initial experiment in order to test if their neuro-physiological responses had changed over time, or were significantly different based on day dependent factors. For this follow-up experiment, subjects were only played their selected Chill segment, and went through a single monitoring cycle. A single authentication attempt was generated for each of the two subjects which was a comparison between their registered responses and their collected responses after a variable amount of time.

\begin{figure*}[t!]
	\begin{subfigure}[t!]{0.3\textwidth}
		\caption{}
		\includegraphics[width=\textwidth]{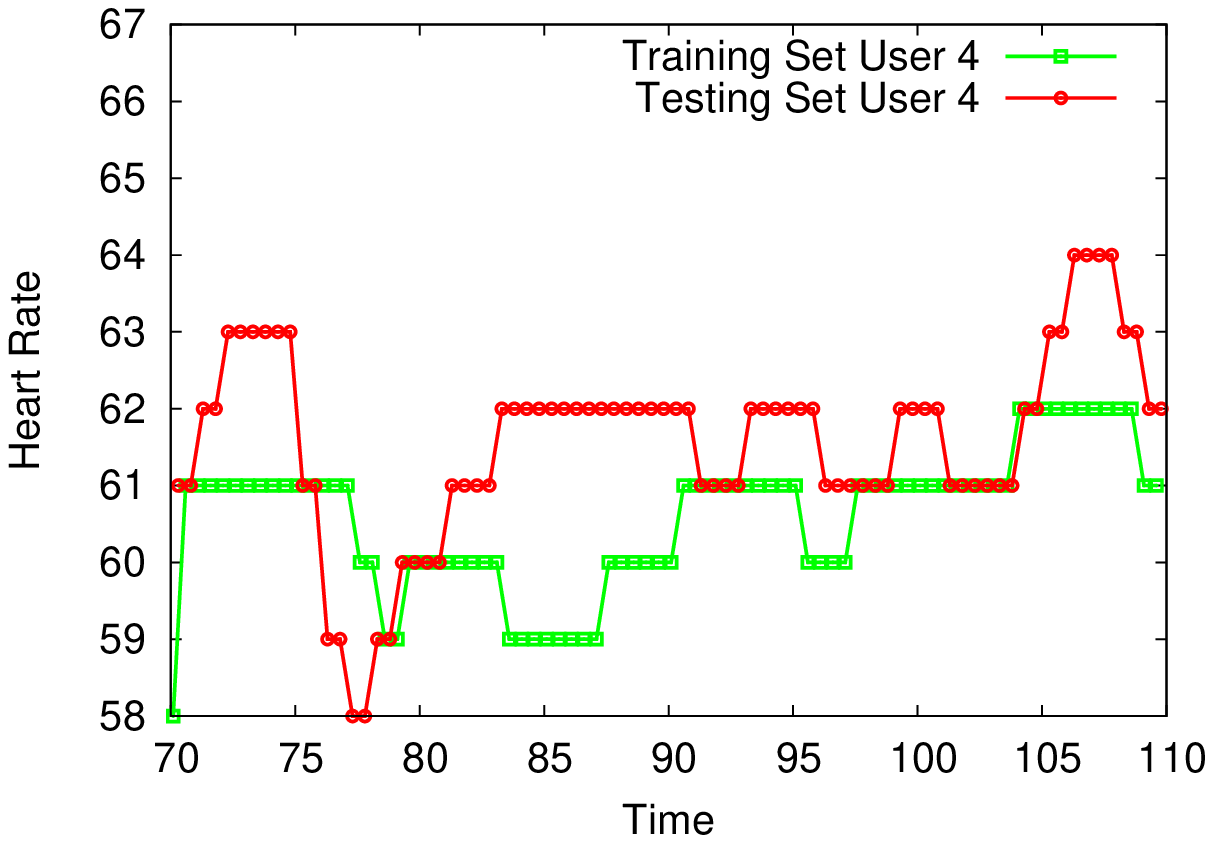}
		\label{fig:multiA}
	\end{subfigure}
	\begin{subfigure}[t!]{0.3\textwidth}
		\caption{}
		\includegraphics[width=\textwidth]{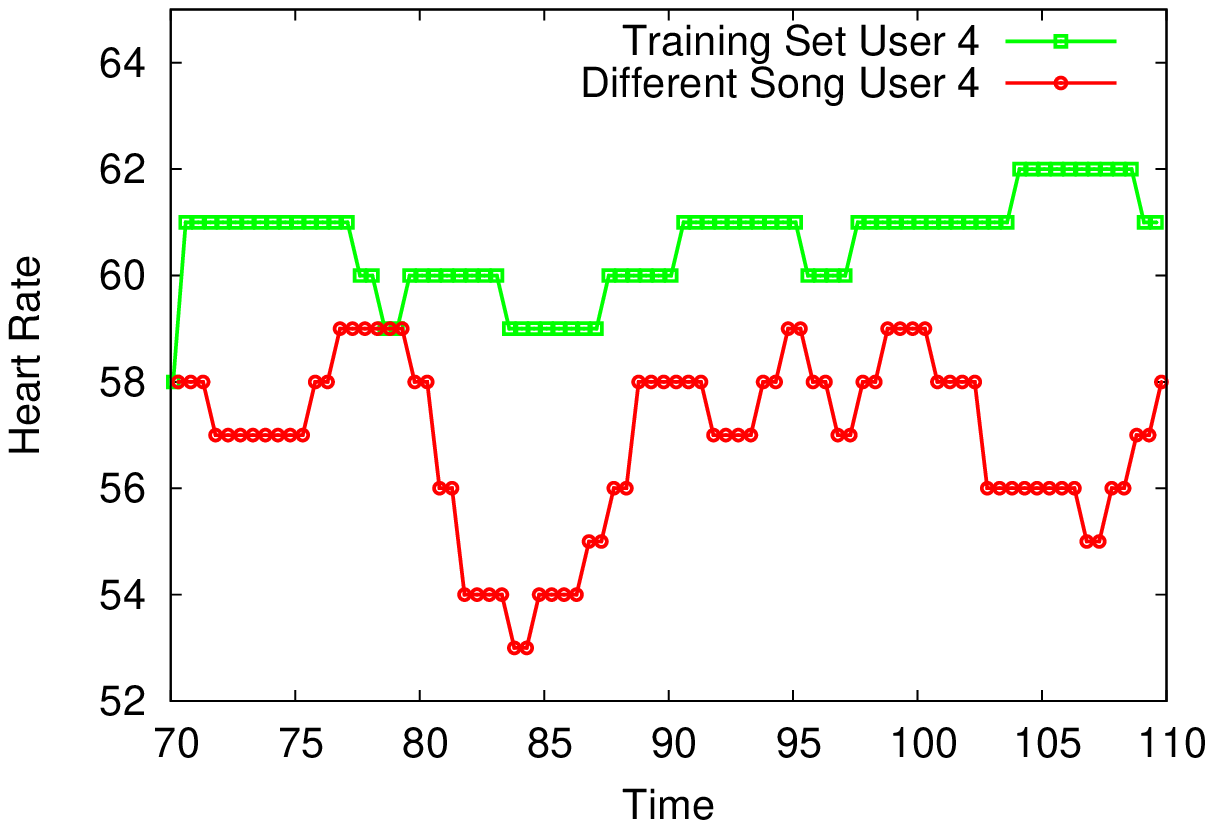}
		\label{fig:multiB}
	\end{subfigure}
	\begin{subfigure}[t!]{0.3\textwidth}
		\caption{}
		\includegraphics[width=\textwidth]{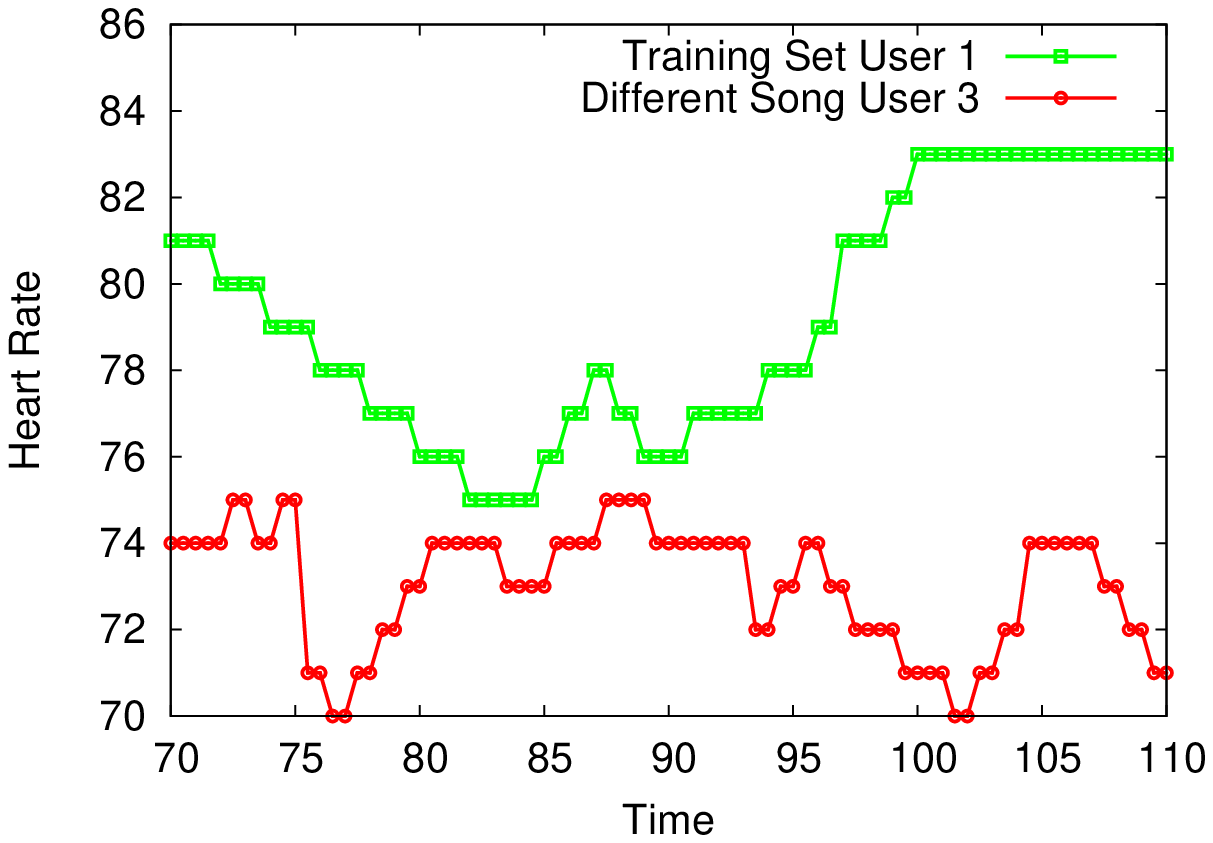}
		\label{fig:multiC}
	\end{subfigure}
	\begin{subfigure}[t!]{0.33\textwidth}
		\caption{}
		\includegraphics[width=\textwidth]{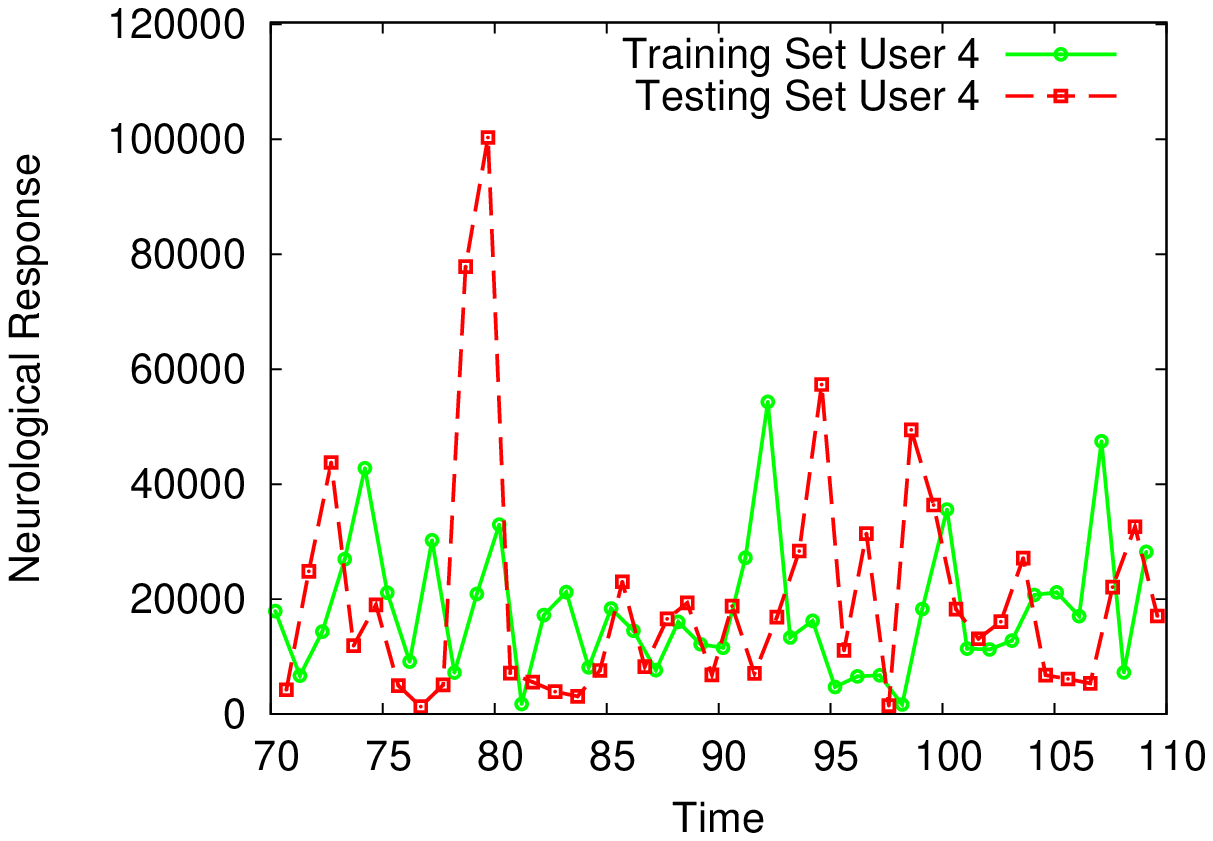}
		\label{fig:multiD}
	\end{subfigure}
	\begin{subfigure}[t!]{0.33\textwidth}
		\caption{}
		\includegraphics[width=\textwidth]{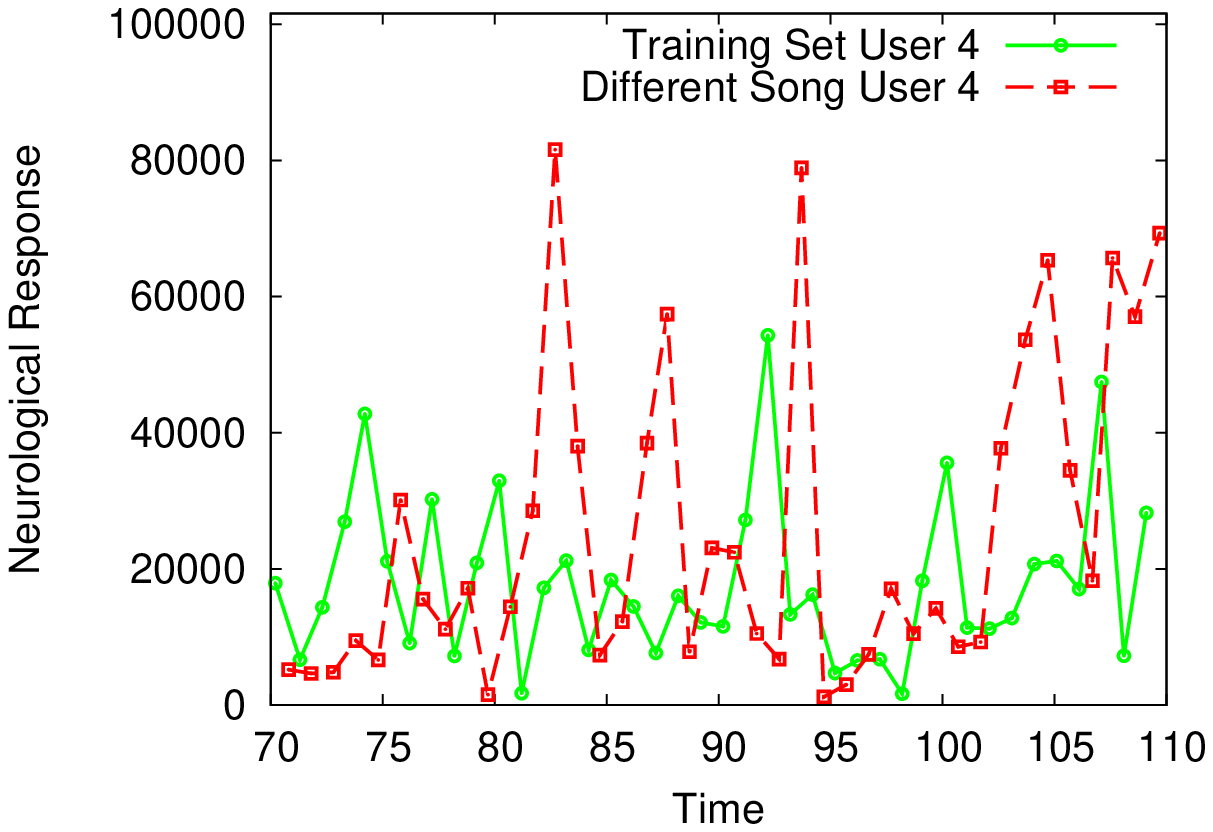}
		\label{fig:multiE}
	\end{subfigure}
	\begin{subfigure}[t!]{0.33\textwidth}
		\caption{}
		\includegraphics[width=\textwidth]{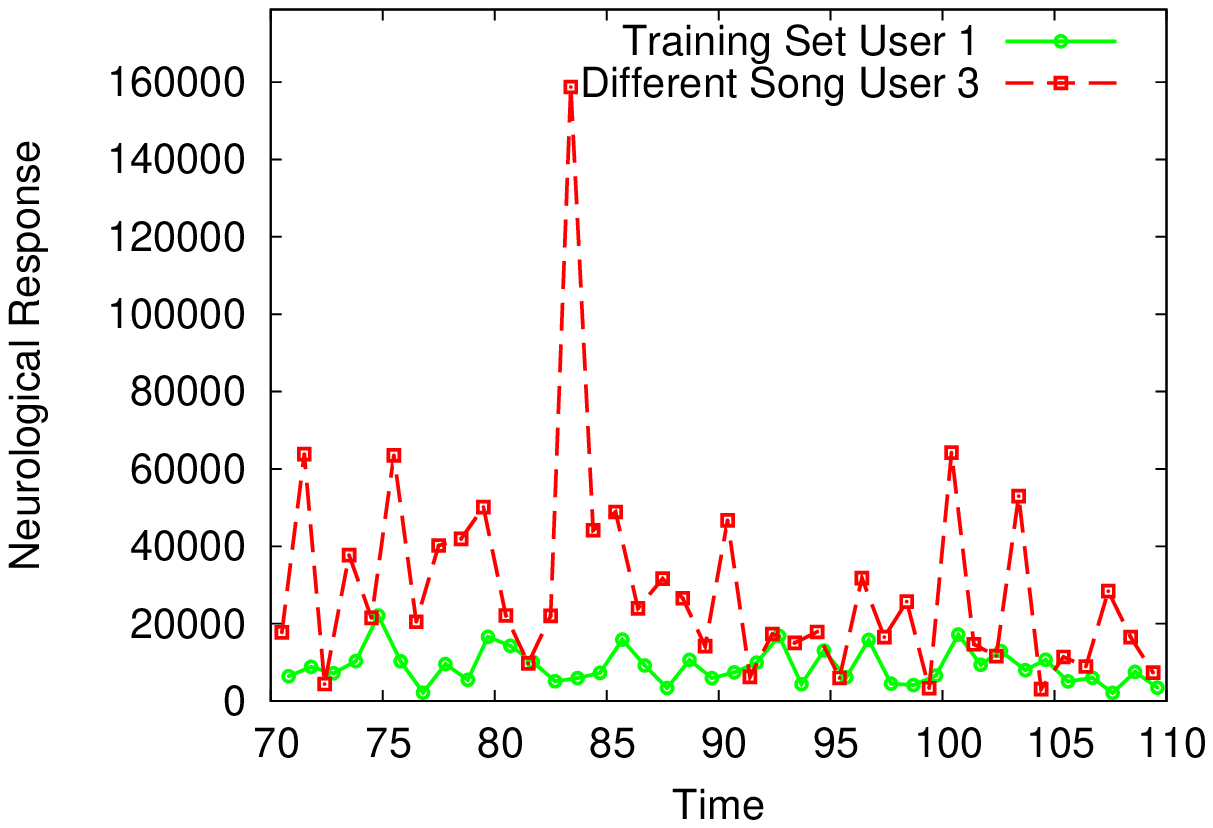}
		\label{fig:multiF}
	\end{subfigure}
	\caption{\textit{The neurological and physiological responses shown above represent authentication attempts collected from the four different categories of comparison. Graphs a) and d) represent authentication attempts consisting of a valid user attempting to authenticate with the registered piece of music. Graphs b) and e) show a valid user attempting to authenticate with a different piece of music than registered. Graphs c) and f) show an invalid user attempting to authenticate with a different piece of music than registered. All neurological responses monitored are Alpha1 brain waves, and all physiological responses are heart rates.}}
	\label{fig:multigraphs}

\end{figure*}

\begin{figure}
	\centering
	\begin{subfigure}[h!]{0.3\textwidth}
		\caption{}
		\includegraphics[width=\textwidth]{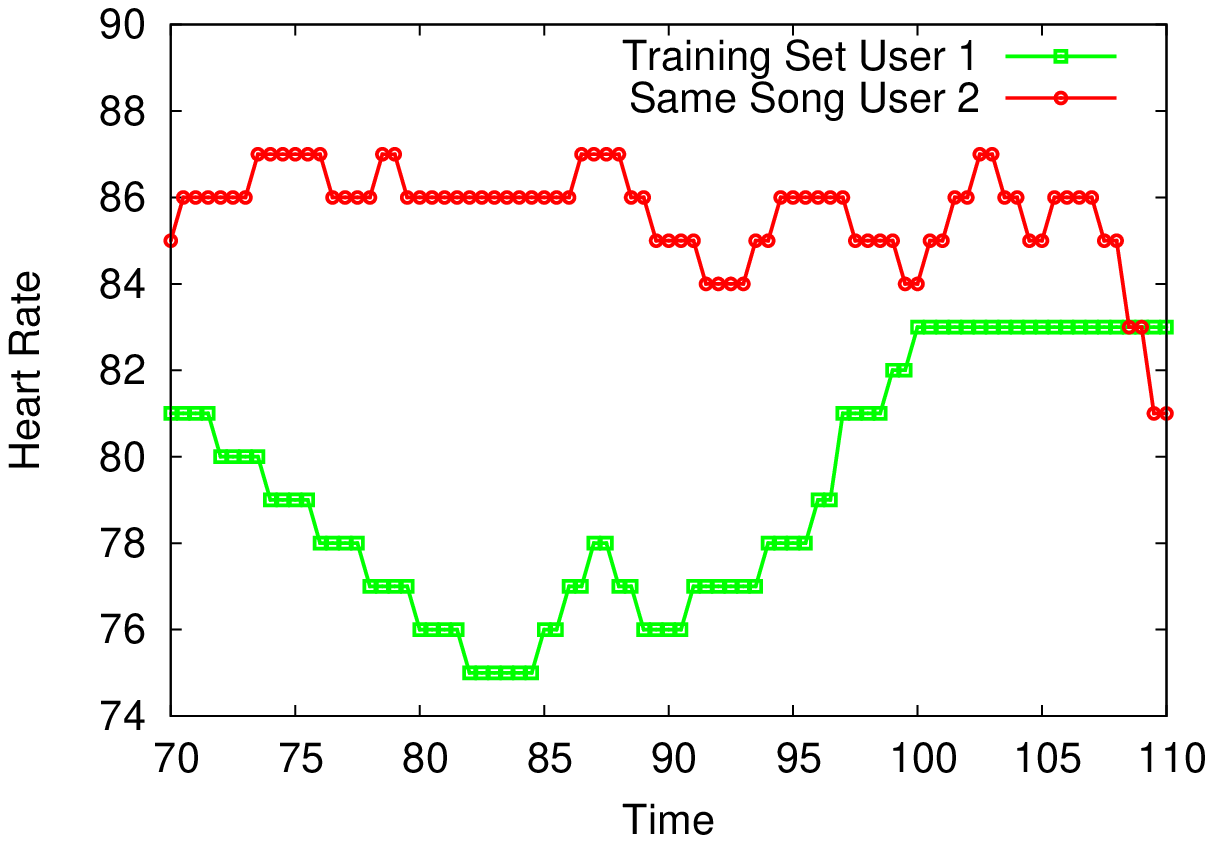}
	\end{subfigure}
	\begin{subfigure}[h!]{0.33\textwidth}
		\caption{}
		\includegraphics[width=\textwidth]{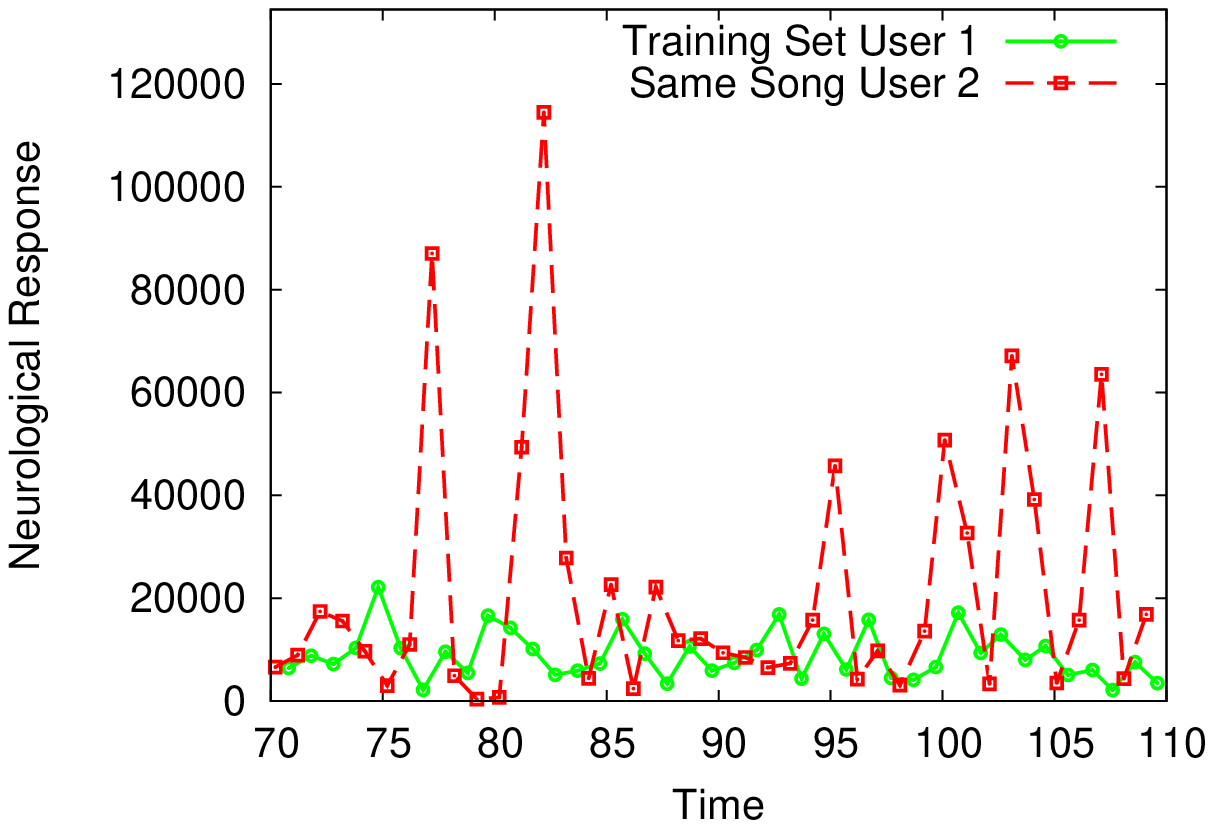}
	\end{subfigure}	
	\caption{\textit{The graphs show an invalid user's physiological (a) and neurological (b) responses as they attempt to authenticate with the same piece of music as registered.}}
	\label{fig:samemusic}
\end{figure}

\section{Experimental Results}
Once neuro-physiological responses were collected from all subjects, we simulated authentication attempts to our prototype CRAS by comparing the neuro-physiological responses registered to each user to all authentication responses collected. Because each subject had one set of registered responses, and four sets of authentication responses (one having the same piece of music as the registered responses), each subject had four authentication responses to be compared to every subject's registered responses (including their own). For five subjects, this meant that 100 possible authentication attempts existed. Out of these attempts, five were acceptable authentication attempts where the user and music were the same for the registered and authentication samples. 15 were authentication attempts where the subject was the same, but not the music. 75 attempts where the subject was different and the music was different, and five attempts where the subject was different, but the music was the same. \\
\indent The first three categories of authentication attempts are shown visually in Figure~\ref{fig:multigraphs}, and the category where the subject was different but the music was the same is shown in Figure~\ref{fig:samemusic} in order to demonstrate our ability to accurately differentiate between authentication attempt categories. We show that authentication attempts where the registered and authenticating subjects are the same have much higher similarity in neuro-physiological responses than cases where the registered and authenticating subjects are different. For neurological responses, only Alpha1 brain waves are shown in order to provide a simplified and consistent representation of neurological responses.

\subsection{Comparing Authentication Attempts}
For each authentication attempt, both the neurological and physiological responses of the samples were compared. All attempts were scored with a coefficient of difference, which was an identifier that was used to show the relative difference in neuro-physiological responses between two collected samples. The coefficient of difference was based on both physiological and neurological responses, and a series of sub-routines that were performed on each set of responses. Different sub-routines were used on physiological and neurological responses, and a coefficient of difference was determined for physiology and neurology separately by averaging the corresponding coefficients of difference from the sub-routines performed. To determine the final coefficient of difference, the physiological and neurological coefficients of difference were summed.  \\
\indent The physiological sub-routines consisted of a categorical test and a penalty test. In the categorical test, the physiological responses from the registration and authentication phase were placed in bins reflecting a range in which they lied between. The bins for the registration and authentication responses were then compared to determine what percentage of responses fell into the same bin at the same time. The penalty test was used to determine the relative difference of the physiological responses at each collected point, and a larger penalty was given for greater difference in response. The total penalty was then calculated to determine if the responses were reasonably similar throughout the authentication attempt. \\
\indent The neurological sub-routines consisted of a penalty test and an average test. The penalty test is similar to the one performed for physiology, however the average test consisted of taking the root-mean squared, geometric mean, harmonic mean, and average of the neurological responses. The more significant the percentile difference of the averages, the higher the coefficient of difference would be. All coefficients of difference determined during the average tests were averaged with the coefficient of difference from the penalty test to determine the neurological coefficient of difference.

\begin{table}[!t]
	\scalebox{.75}{
		\centering
		\begin{tabular}{|c|c|c|}
			\hline
			\textit{\textbf{Authentication Type}}                                                       & \textit{\textbf{Number Passed / Total}} & \textit{\textbf{Percentage}} \\ \hline
			\begin{tabular}[c]{@{}c@{}}Valid Attempt\\ \textit{(Same User Registered Music)}\end{tabular}        & 5/5                                     & 100.00\%                     \\ \hline
			\begin{tabular}[c]{@{}c@{}}Cross Attempt 1\\ \textit{(Same User Different Music)}\end{tabular}       & 10/15                                   & 66.67\%                      \\ \hline
			\begin{tabular}[c]{@{}c@{}}Cross Attempt 2\\ \textit{(Different User Registered Music)}\end{tabular} & 0/5                                     & 0.00\%                       \\ \hline
			\begin{tabular}[c]{@{}c@{}}Invalid Attempt\\ \textit{(Different User Different Music)}\end{tabular}  & 7/75                                    & 9.33\%                       \\ \hline
		\end{tabular}
	}
	\caption{\textit{The accuracy of the system can be shown by how many samples of each category passed a simple authentication test. By passing all authentication attempts with a coefficient of difference less than or equal to 2.0, and failing all others we show how many samples from each category can pass this authentication test.}}
	\label{tab:percentlist}
\end{table}

\subsection{The Results}
Of the 100 authentication attempts collected, all 5/5 valid authentication attempts passed the authentication test, 10/15 of the same subject attempts passed the authentication test (due to the similarity in baseline measurement as discussed in detail in Section~\ref{sec:sameuser}), 0/5 of the same music attempts passed the authentication test, and 7/75 of the unacceptable attempts passed the authentication test. The simple authentication test we used in this experiment was to define a coefficient of difference of 2.0 or less as passing, and greater than 2.0 as failing. We chose 2.0 because the final coefficient of difference is calculated by summing the coefficient of differences for physiology and neurology, and we expect in a valid sample for each of those coefficients of difference to be below or equal to 1.0. \\
\index We used the number of authentication attempts that passed our authentication test as a metric for percent error in order to determine the effectiveness of our system. As shown in Table~\ref{tab:percentlist}, valid attempts all passed the authentication test, so false negatives were not present in our simple authentication test. However, as shown in the data for invalid attempts, 7/75 samples passed the authentication test, giving 9.33\% false positive results. When looked at as a whole, the same music samples can also be attributed to invalid attempts, and we mainly separated them in order to show that playing the same music to a different user did not give an invalid user a better chance of authenticating. If we consider same music attempts as invalid attempts, then our false positive rate is reduced to 7/80, or 8.75\%. \\
\indent Because same music attempts are a subset of invalid attempts as shown above, we can state that the ability of our system to accurately identify authentication attempts is 78/85, or 91.76\%. We calculate this by accurately classifying 5/5 valid authentication attempts and 73/80 invalid authentication attempts. Although same user attempts were also collected, they are not included in this metric because they are not relevant in identifying the accuracy of the system in practice. One thing we did not check when comparing authentication attempts was if both sets of neuro-physiological responses being compared had experienced the Chill effect. The registered responses were guaranteed to have a Chill response, and all of the valid authentication attempts were checked to make sure they contained a Chill response, however none of the other attempts were checked in this way. There are also several other factors in the neuro-physiological responses we could have used to reduce our false positive rate, such as checking the baseline neuro-physiological responses are imposing automatic fails if certain anomalous traits were identified. We chose to not to impose these restrictions during our experiment as our goal was to compare the consistency and uniqueness of neuro-physiological responses, and not to simulate a complete authentication system.\\
\indent Visualizations of the neuro-physiological responses for the different categories of authentication attempts are also available in Figure~\ref{fig:multigraphs} and Figure~\ref{fig:samemusic}. We can see that the neuro-physiological responses that are most similar are those from Figure~\ref{fig:multiA} and Figure~\ref{fig:multiD} which correspond to the physiological and neurological responses respectively during an authentication attempt where the subject is the same as the one registered, and they are listening to the same piece of music they registered. The responses in Figure~\ref{fig:multiB} and Figure~\ref{fig:multiE} correspond to the case where the subject is the same as registered, but they are listening to a different piece of music. Although these responses are more similar to each other than cases where the subject is different, they are clearly less similar than the case where the subject and the music being played are the same. Figure~\ref{fig:multiC} and Figure~\ref{fig:multiF} show the case a different subject is trying to authenticate with a different piece of music than registered, and Figure~\ref{fig:samemusic} shows the same case but with the same piece of music as registered. The responses from both of these cases are not easily discernible from each other, and they both result in neuro-physiological responses which are clearly different than those registered into the system.

\begin{figure}[!t]
	\includegraphics[width=.5\textwidth]{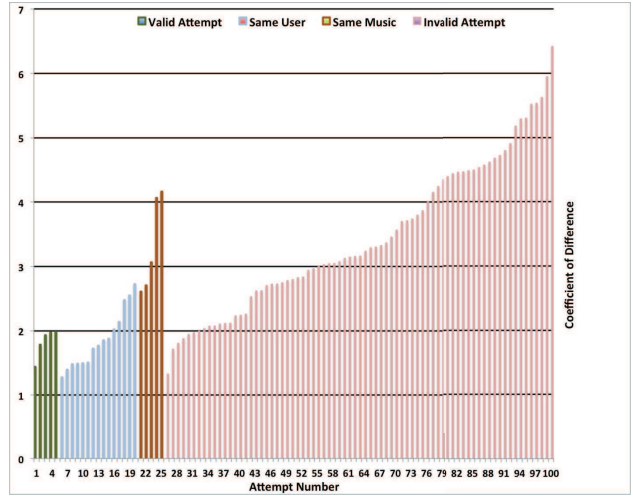}
	\caption{\textit{The authentication attempts were separated into four categories (from left to right): (leftmost/green) valid authentication attempts, (blue) same user attempts, (orange) same music attempts, and (rightmost/red) invalid authentication attempts. Each subcategory is plotted in ascending order, with the x-axis showing an arbitrary attempt number and the y axis showing the coefficient of difference determined by summing the coefficients for neurological and physiological responses.}}
	\label{fig:responsegraph}
	
\end{figure}

\subsubsection{Same User Case} \label{sec:sameuser}
The first thing to note is that the same user has an initial advantage in the system because their baseline responses will be similar in all authentication attempts, regardless of the music they are listening to. It is also important to note that although Dopamine releases due to music affect different subjects uniquely, the Chill responses to multiple pieces of music from a single subject may demonstrate similar neuro-physiological responses. Because of this, even though a subject is experiencing a Dopamine release due to a different piece of music than they registered, their neuro-physiological responses may still change in a way similar to if they had been listening to their registered piece of music. \\
\indent Although we consider the same user case for experimental purposes, in practice it is negligible. This is because whenever the valid user attempts to authenticate into the authentication system, they will always listen to music they have already registered, so our experimental same user case cannot exist. The case of a different user listening to a different piece of music is similar, but it is possible that an attacker may choose to listen to a different piece of music in the hopes of replicating the same responses the valid user had to the registered music. From our results this kind of attack seems to be ineffective, however because of this possibility we do not also discount that case. It is also worth mentioning that a valid user could attempt to listen to an incorrect piece of music when authenticating in order to try and deceive the stress detection, however due to Dopamine releases due to stress and Chill music occurring in different locations in the brain, this attack does not seem to be feasible. \\

%

\begin{table}[t]
	\centering
	\begin{tabular}{|c|c|c|}
		\hline
		\multicolumn{1}{|l|}{\textbf{Subject Number}} & \multicolumn{1}{l|}{\textbf{Time Elapsed}} & \multicolumn{1}{l|}{\textbf{\% Difference}} \\ \hline
		3                                           & 1 Month                                   & 6.73\%                                           \\ \hline
		5                                           & 1 Week                                    & 3.41\% \\ \hline                
	\end{tabular}
	\caption{\textit{Two subjects were selected to return and attempt to authenticate into our CRAS after different periods of time. The percentile difference of their neuro-physiological responses from the multiple occasions they attempted to authenticate are given.}}
	\label{tab:TimeDecay}
\end{table}

\subsection{Day Dependence and Time Decay}
\begin{figure}[t!]
	\centering
	\begin{subfigure}[t!]{0.3\textwidth}
		\caption{}
		\includegraphics[width=\textwidth]{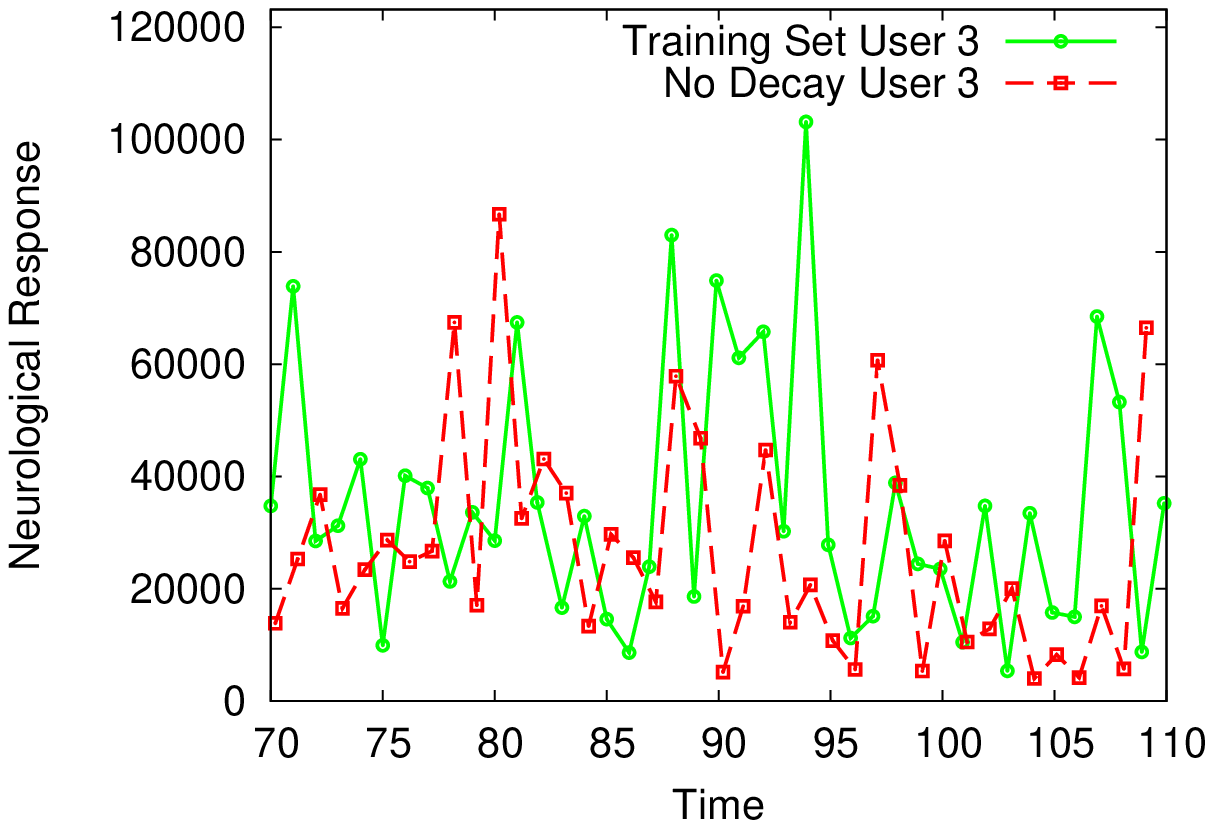}
	\end{subfigure}
	\begin{subfigure}[t!]{0.3\textwidth}
		\caption{}
		\includegraphics[width=\textwidth]{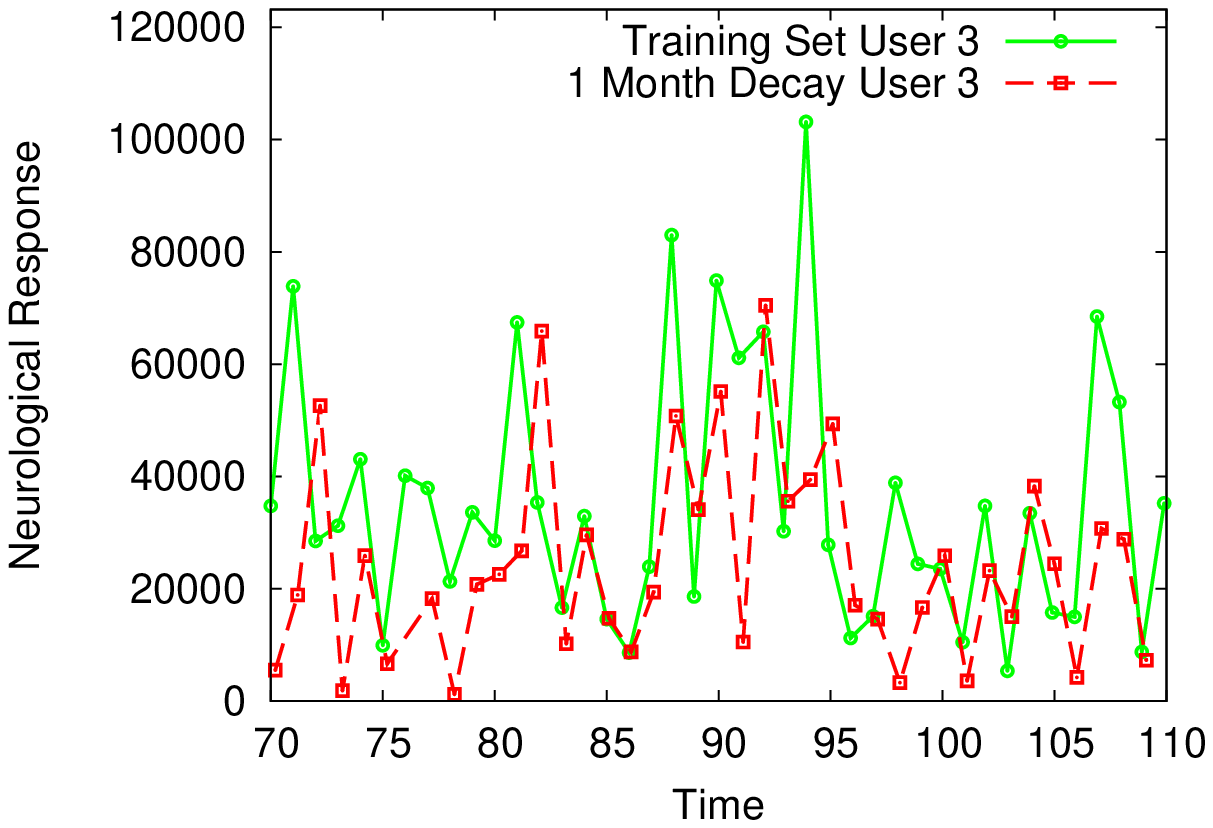}
	\end{subfigure}
	\caption{\textit{The neurological responses collected from a valid authentication attempt during the preliminary experiment as well as those collected a month after the experiment are shown above. The neurological responses shown are Alpha1 brain waves as in the previous example.}}
	\label{fig:TimeDecay}
\end{figure}

\indent Because day dependence and time decay were not covered by our preliminary experiment, we invited two of our subjects to come back on separate occasions of variable time after their initial registration into the CRAS, and conduct authentication again in order to measure day dependence and time decay. Table \ref{tab:TimeDecay} shows the time elapsed between the preliminary experiment and the authentication attempt collected afterwards, and the percentile difference of the neuro-physiological on those two occasions. A visual of the change in neurological responses over time is shown in Figure~\ref{fig:TimeDecay}.\\
\indent Overall we show that across the two subjects monitored, there is a 3-7\% difference in their neuro-physiological responses when attempting to authenticate into the system after an extended period of time after their original registration. This change over time is not extreme, and does not seem to significantly affect the subjects ability to accurately authenticate into the system. Due to our music selection and registration process, it seems that the effects of day dependence and time decay were successfully mitigated.

\section{Discussion}
Our experiments have shown that it is possible to classify neuro-physiological responses during the stimulation of Chill music based on their similarity to registered responses, but they have not shown how we have addressed all concerns necessary to make a CRAS. Also, because our experimental setting does not perfectly mirror the real world, some things that occur during experimentation may not occur in the real world. These differences that occurred during the data collection process, the issues of day dependence and time decay, the uses and constraints of Chill segments, and the ability of the proposed system to defend against coercion attacks are all important aspects of the proposed CRAS which were not covered in the preliminary experiment, but are discussed below in detail.

\subsection{Data Collection}
Unlike the registration phase, our simulation of the authentication phase during our experiment does not perfectly reflect a real world use-case. The reason is that it is unlikely that a user will have listened to their registered music directly before attempting to authenticate, and it is also unlikely that they will have been listening to music continuously before authentication. However, in experimentation both of these cases are common. To mitigate inconsistency, we had a consistent order in which we played subjects pieces of music, so that each subject listened to each category of music in relatively the same order (some exceptions were made in the case of equipment malfunction or failure to select Chill music). Still, before conducting the authentication phase for their selected Chill music each subject had at least listened to that piece of music three times. \\
\indent Because of this difference from the real world, it is possible that the responses that we collected during our experimentation are more stale. A stale response can be defined as a response collected after doing the same task multiple times in quick succession, where the response becomes less potent over subsequent iterations. Responses that are stale are different than responses affected by time decay in an important way. Responses that are stale become less potent because they are done in quick succession, and return to normal if the task is performed again after a long enough break. However, results that become less potent due to time decay decrease because over the course of time the user's responses fundamentally change, and will most likely not revert to what they were previously. The discrepancy of our results due to stale responses is likely to affect our results negatively instead of positively, so in theory our experimental results should be slightly worse than if the system was used in practice.


\subsection{Music Anhedonia}
Another topic that was not covered in our experiment was music anhedonia, which refers to people who do not respond to music~\cite{satoh2011musical}. This concept has been studied in previous work, however is still under research and is not completely understood. For the purposes of our work, we did not encounter any subjects who failed to respond to music, however we consider it a possibility and plan to look into the scenario in future work.

\subsection{Coercion Attacks}
Although our proposed authentication system is meant to be coercion resistant, no where in our experimentation did we explicitly introduce coercion to our subjects. One reason we did not do this is because it is unethical to threaten test subjects in order to verify that our system is fully coercion resistant, as it could leave subjects with permanent physical or psychological damage. Also, simulating a coerced state by inducing extreme stress in a subject could possibly result in similar damage, which is undesirable and we felt was largely unnecessary. Above this, it is extremely difficult to receive IRB approval for this kind of test outside a medical setting. Due to all of these reasons, we chose to use supporting evidence on the effects of coercion on neuro-physiological responses instead of collecting our own. \\
\indent Previous work shows us a clear connection between Dopamine releases and stress, and defends our claim that a Dopamine release caused by a Chill response would be significantly altered during extreme forms of stress. This gives us a guarantee that during a coercion attack, neuro-physiological responses will be too deviated from the norm to be identified as acceptable authentication materials~\cite{pani2000role}.

\section{Related Work}
Our specific implementation of a CRAS using neuro-physiological responses to Chill music has not been attempted previously, so there is no previous work directly linked to it. However, there has been a number of previous work conducted on various components of our research, such as coercion resistant authentication systems, the Chill effect, neuro-physiological responses to stress, and building authentication systems based on neuro-physiological responses. 

\subsection{Coercion Resistant Authentication}
As discussed previously there are two kinds of coercion attacks, those that attempt to force the victim to give the attacker their authentication materials, and those that attempt to force the victim to authenticate into the system and give the attacker access. The prevention techniques which have been highlighted in previous work focus on making these attacks infeasible by either preventing authentication materials from being transferable or by identifying when an authenticating user is being coerced. \\
\indent The prevention technique that has been suggested to combat the first form of coercion attack is the use of implicitly learned tasks as authentication materials~\cite{bojinov2012neuroscience}. By having the authentication system require an implicitly learned task to be performed, there is a guarantee that the only users able to access the system are those who have been previously trained by the system to perform the authentication task. This method adequately satisfies the requirements to prevent the first form of coercion attack, because even under the circumstance of coercion a victim would be unable to transfer their authentication materials to an attacker. Although this system effectively addresses the first form of coercion attack, it does not address the second as their is no guarantee that a coerced user will be unable to perform the required authentication task. \\
\indent To combat the second form of coercion attack it is necessary to have some mechanism of identifying when a user is coerced, so that authentication can be denied even though a valid user is attempting to authenticate. Generally this would require some form of biometric identifier that would be able to alert the system if the authenticating user was under significant stress (likely due to coercion). A suitable biometric that has been used in previous work is skin conductance, because it responds quickly to stress and isn't generally affected by many other factors~\cite{gupta2010fighting}. The drawback of using a biometric that is available during a baseline state is that it is susceptible to theft, and an attacker can possibly steal the value associated with the biometric while the victim is unaware (for example, while sleeping). For this reason, this form of coercion resistance is also only able to defend against one of the aforementioned coercion attacks and not both.

\subsection{Music Based Authentication Systems}
Prior to our own work, music based authentication systems have been suggested~\cite{gibson2009musipass}, but previous attempts were not based on using Chill music. A significant difference in our work is our focus on the Chill effect, and our ability to authenticate consistently is reliant on the reproducible nature of neuro-physiological responses stimulated by a Dopamine release. The Chill effect, and corresponding neuro-physiological responses during the Dopamine release associated with it, have been studied heavily in previous work~\cite{grewe2009chill,salimpoor2011anatomically}. 

\subsection{Neuro-Physiological Authentication}
Biometric authentication methods offer many benefits over alphanumeric passwords, and widely studied in the literature. In the domain of biometric authentication systems, the research most relevant to our work are those that focus on using heart rate or neurology to develop authentication systems, and those that detail the change of these responses due to stress. \\
\indent There have been several studies highlighting the feasibility of accurately monitoring heart rate~\cite{camara2015non}, the consistency of heart rate as a biometric~\cite{thayer2012meta}, and the ability to authenticate a user by using their heart rate as an authenticating factor~\cite{da2013finger}. These studies are important because they highlight the ability of user's heart rate to be used as an effective authenticating factor. On the side of neurology, there were also several studies concerning developing authentication systems based on neurological responses~\cite{campisi2014brain,jolfaei2013feasibility,maiorana2015cognitive}. Because Dopamine is released in the brain during a Chill effect, it is important that we are able to use a user's neurological responses as authentication material, so the existence of these methods were very important. \\
\indent Coercion is almost always accompanied by high levels of stress, so being able to detect if a user is under significant stress is essential for a CRAS to perform desirably. There have been multiple studies that have shown how stress affects a user's physiological~\cite{everly2002anatomy,gupta2012coercion,thayer2012meta} and neurological~\cite{pani2000role} responses. Work has also been done to identify the location in the brain a Dopamine release due to stress occurs~\cite{StressDopamine}. Leveraging the results of these researches, we are easily able to identify when coercion occurs by monitoring the change of neuro-physiological responses, and know that the reaction is different from the Dopamine release to Chill music because they occur in different parts of the brain.

%
%

\section{Conclusion}
We have experimentally validated our proposed Chill music based CRAS by attempting to answer four fundamental questions about a user's neuro-physiological responses to Chill music: are they consistent, do they change over time, are they similar to other user's responses for that music, and are they similar to other user's responses for any other music. \\
\indent The consistency of a subject's neuro-physiological responses to Chill music were shown by our accurate classification of authentication attempts where the authenticating user was the same as the registered user, and was listening to the same Chill music they registered with. When analyzing the consistency of these neuro-physiological responses over time, we found that although some level of decay of responses was present, the decay was not significant enough to be cause false negatives. \\
\indent Our results support the feasibility of a Chill music based CRAS, however the usability issues were not considered in the scope of this work. With the increasing popularity of wearable technologies, such an authentication system based on neuro-physiological responses becomes both possible and practical. Whether it is to be used as a second factor for authentication, or a one-time mechanism to be supported by continuous authentication, there are several practical applications for using our proposed authentication system for bolstering previously existing authentication systems, or being used stand-alone. \\
\indent Although our experiments do not boast perfect classifications of authentication attempts, our above 90\% accuracy highlights the ability of our prototype system. By playing multiple pieces of music to a user instead of a single piece, or using our system as a second factor, the non-perfect accuracy can be mitigated and the result can be an extremely powerful, and coercion resistant, authentication technique. Not only are there applications for such research in government and other high-security facilities, but also normal users could benefit from this authentication method by giving them authentication materials that they could not lose or forget, and that could easily be retrieved within a minute of listening to a registered piece of music. \\
\indent In the future we wish to build on our proposed system in order to make it more accurate, and more convenient. One such addition would be a Chill music selector, which could help a user select music that was likely to elicit a Chill response based on their personal preference in music, and their responses to previous kinds of music. We would also like to explore techniques to combine segments of multiple pieces of music in a way that could create multiple Chill segments per piece of music, and provide an even stronger authentication. We are also interested in exploring if it is possible to profile malicious intent through such an authentication technique, in order to prevent possible insider attacks before they occur. 
{\footnotesize \bibliographystyle{acm}
\bibliography{bibliography}}
%
%

\end{document}